# The Origin of Chiral Life

by
Vlado Valković and Jasmina Obhodas
Institute Ruder Boskovic
Zagreb, Croatia

**Abstract:**

The phenomenon of life is discussed within a framework of its origin as defined by four hypotheses. The 1. hypothesis says: Life, as we know, is (H-C-N-O) based and relies on the number of bulk (Na-Mg-P-S-Cl-K-Ca) and trace elements (Cr-Mn-Fe-Co-Ni-Cu-Zn-Se-Mo-I-W, and possibly Li-B-F-Si-V-As). It originated when the element abundance curve of the living matter and of the Universe, coincided. The 2. hypothesis is: Life originated in an interstellar molecular clouds with the critical role of dust particles. The 3. hypothesis arises from the 1. and states: Because of the Universe aging, life originated only once. The dust forming planetary system and stars already contained an excess of L-type amino acids and D-type sugars, therefore, the emerging life on any planet had to be chiral. Consequently, the 4. hypothesis has been formed: Chirality is a *sine qua non* condition for the emergence of life. The arguments supporting these hypotheses are put forward based on numerous astrophysical observations and physics laws.

## 1. Introduction

According to the present-day thinking about the phenomenon of life (Barrow et al. 2012), life is a spatially distinct, highly organized network of chemical reactions characterized by a set of remarkable properties that enable it to replicate itself and to adapt to changes in its environment. However, this actually describes how life on Earth sustains rather than how it originated. Here we shall present different views about the origin of life problem and suggest some possible future activities which might prove these assumptions.

The chemical evolution of galaxies results in the changes of chemical composition (relative abundances of chemical elements) of stars, interstellar gas, and dust. The increase of chemical element abundances with time (elements heavier than Li) provides a clock for galactic aging. The older the galaxy, the more heavy elements contain. On the other hand, looking at the living matter (the one we know, on the planet Earth) we see that life needs only some chemical elements for its existence. All organisms use trace elements and provide proteins with unique coordination and catalytic and electron transfer properties.

Therefore, to shed some light on the problems where and when life originated, we are putting forward **the 1. hypothesis**. It says: *Life, as we know, is (H-C-N-O) based and relies on the number of bulk (Na-Mg-P-S-Cl-K-Ca) and trace elements (Cr-Mn-Fe-Co-Ni-Cu-Zn-Se-Mo-I-W, and possibly Li-B-F-Si-V-As). It originated when the element abundance curve of the living matter and the one of the Universe, coincided.* This coincidence occurring at a particular redshift could indicate the phase of the Universe when life originated, $T_{origin}$.

Interstellar molecular clouds and circumstellar envelopes are places of complex molecular synthesis (Ehrenfreund and Charnley 2000; Kwok 2004; van Dishoeck 2014). In addition to gas, interstellar material also contains small micron-sized particles. Gas-phase and gas-grain interactions result in the formation of complex molecules. Surface catalysis on solid particles enables molecule formation and chemical pathways that cannot proceed in the gas phase because of reaction barriers (Scorei 2012). A high number of molecules that are used in contemporary biochemistry on the Earth are present in the interstellar medium on surfaces of comets, asteroids, meteorites, and interplanetary dust particles. Therefore, o*ur* **2. hypothesis** *is that life originated in an interstellar molecular clouds with the critical role of dust particles in their environment.*

The Solar system we live in is the result of the gravitational collapse of a small part of a giant molecular cloud. Assuming life originated long before the origin of Earth, it could be hypothesized that life was brought to Earth in the process of its formation. Here we must consider the role of molecular cluds cosmic dust mineral grains in this process. It is reasonable to assume that life will not depend on resources that are scarce in its environment (McClendon 1976). Yet, the chemical element boron, which is depleted in the Solar system but not in the interstellar medium, played an essential role in the process of life forming since its primary purpose has been to provide thermal and chemical stability in hostile environments (Scorei 2012). Rafelski et al. (2012 and 2014) showed the rapid decline in the metallicity (in astrophysics, all elements heavier then hydrogen and helium are metals) of damped Lyα systems at $z > 4.7$. They presented chemical abundance measurements for 47 damped Ly-α systems (DLAs), 30 at $z > 4$, observed with the Echellette Spectrograph and Imager, and the High-Resolution Echelle Spectrometer on the Keck telescopes. It looks that the properties of the dust in DLAs would not satisfy the requirements of essential trace element signature of Last Universal Common Ancestor (LUCA; concentration values ratios - Cr: Mn: Fe: Co: Ni: Cu: Zn: Se: Mo: I: W) for $z > 4.7$ (i.e., before $\approx -12.6 \times 10^9$ years or $\approx 1.1 \times 10^9$ years after the Big Bang). At some point in time, the chemical evolution of the Universe resulted in the elemental abundance curve coinciding with LUCA elemental signature. For any other time, $T \neq T_{origin}$, no additional such events could occur. The time when the coincidence of element abundance curves for the Universe and LUCA occurred could not be older than $T \approx -12.6 \times 10^9$ years, a lower limit for $T_{origin}$, because the Universe was chemically too young to produce elements in the right elemental ratios as required by LUCA. Following from the 1. and 2. hypothesis, the **3. hypothesis** can be formed, that is, *because of the Universe aging, life originated only once.* It is proposed to look into a redshift region $z = 0.5 – 4.7$ (approximately $T = -5.2 \times 10^9$ years to $-12.6 \times 10^9$ years). DLA systems in this region have been studied, and one should use this data to study the evolution of chemical element abundances in galaxies to determine the time when the universal element abundance curve coincided with the element abundance curve of LUCA. The genetic code has transmitted the characteristic properties of the latter while the universe element abundance curve changed as the galaxies aged.

The dust that formed our Solar system already contained L-type amino acids and D-type sugars in excess, therefore, the life on planet Earth, irrespective of its origin, had to be chiral. This can be generalized to planets in any other solar system. At this point we can put forward our **4. hypothesis:** *Chirality is a sine qua non condition for the emergence of life.*

## 2. Chemical Evolution of the Universe

The cosmic background radiation, the remnant of the first event, the Big Bang, is substantially uniformly distributed in the Universe (COBE 2015). The measured cosmic microwave background spectrum is almost perfect blackbody radiation with a temperature of 2.725 +/- 0.002 K. This measurement matches the predictions of the hot Big Bang theory exceptionally well. It indicates that nearly all of the radiant energy of the Universe was released immediately after the Big Bang. Contrary to radiation, the distribution of matter is highly non-uniform. Galaxies occupy $10^{-7}$ of the volume of the Universe but contain most of the known matter. The sharp aggregation of matter means that the chemistry is occurring within galaxies. The small particles in the space between the stars have surfaces convenient for the production of many simple and complex molecules. The existence of molecules in the low-density matter that exists between the stars has been first reported by Adams (1949), and the list of the observed molecules (http://www.astro.uni-koeln.de/cdms/molecules) has been increasing since then.

The list contains a new addition, $HeH^+$, as described in the paper by Güsten et al., 2019, found in the planetary nebula NGC 7027. This molecule is the first type ever formed in the Universe: around $10^5$ years after the Big Bang. Its destruction created a path to the formation of molecular hydrogen through the series of reactions:

$$He^+ + H \rightarrow HeH^+ + h\nu$$
$$HeH^+ + e \rightarrow He + H$$
$$HeH^+ + H \rightarrow He + H_2^+$$

Molecular hydrogen is the molecule primarily responsible for the formation of the first stars. Dust affects star formation through molecular gas formation, see Fig.1.

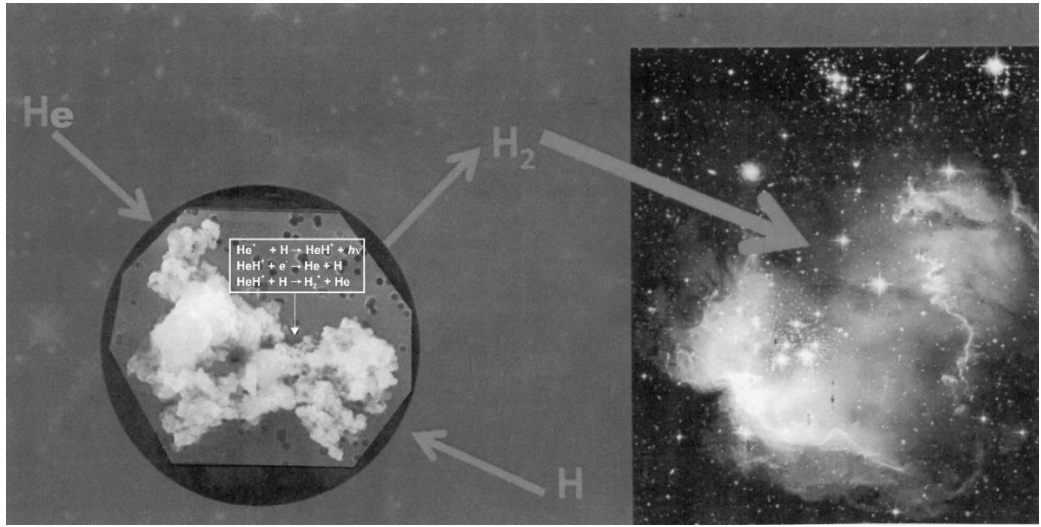

Fig. 1: Schematics of molecules formation on dust particles.

Dust particles have an enormous contribution to the energy budget of a galaxy. In our galaxy, dust presents less than 1% of mass but contributes to about 30% of luminosity (Rémy-Ruyer et al. 2014). As galaxies evolve, their interstellar medium becomes continually enriched with metals, and this metal enrichment influences the subsequent star formation.

The main characteristics of interstellar solid particles are their size, shape, chemical composition, and amount. The most obvious consequence of the dust presence is the extraordinary blocking of the light of the stars. As seen from the average interstellar extinction curve, the light in red is reduced less than the light in the blue and the ultraviolet. From the shape of the extinction curve, one can deduce that the particles responsible for the visual extinction are about 0.1 µm in size (a mean radius), and those responsible for the far-ultraviolet disappearance are 10–100 times smaller (Li and Greenberg, 1997). The principal molecular ingredients of the interstellar dust are obtained by studying the way they absorb or emit infrared radiation. Luckily, the so-called fingerprint region of the infrared spectrum from 2.5 to 25 µm, where most vibrations of the molecular groups containing C, N, O, and H occur, is accessible with modern telescopes.

Cosmic dust grains with a carbonaceous and siliceous composition (Draine 2003) represent the most pristine starting material for planetary systems and provide a surface for key astrochemical reactions. Dust grains in cold dense astrophysical environments, such as dense interstellar clouds, protostellar envelopes, and planet-forming disks beyond the snow line, are typically considered to be mixtures of dust particles with molecular ices. Water is the main constituent of these ices, accounting for more than 60% of the ice in most lines of sight (Whittet 2003). Ices are believed to cover the surface of a dust core and/or to be physically mixed with dust.

Molecular clouds are characterized by low temperatures, on the order of 10 K. At these temperatures, there is insufficient energy for collisions to overcome any activation barrier to reaction. Interstellar gas clouds also have extremely low densities, which have drastic consequences for the collision frequency, and therefore the number of opportunities for chemistry to occur. The number of collisions is small, even in the densest regions; with densities of $10^6$ cm$^{-3}$, collision rates are around $5 \times 10^{-4}$ s$^{-1}$, approximately one collision every half an hour. In less dense regions, atoms and molecules may go for many weeks, or even longer, between collisions. Chemistry, therefore, occurs at a prolonged rate in interstellar space compared to the timescales on Earth. The giant molecular clouds last for around 10-100 x$10^6$ years before they are dissipated by heat and stellar winds from stars forming within them. There is plenty of time for some quiet complex chemistry to occur at a rather slow rate.

The very low collision frequency has significant consequences for the types of molecules that may form in interstellar space. Terrestrial concepts of molecular stability do not apply in this extremely non-reactive environment. Carbon does not need to have four bonds; in fact, there are many subvalent species, radicals, molecular ions, and energetic isomers amongst the molecules observed in interstellar gas clouds. Carbon-containing compounds tend to be highly unsaturated, with many double and triple bonds, and few branched chains.

Chemistry can also occur on the surface of dust grains. Surface‐catalyzed reactions of this type turn out to be very important in the interstellar medium. Photoionization is a widespread process near stars. However, the high density of hydrogen and dust grains in molecular clouds prevents visible and UV light from penetrating very far. For this reason, molecular clouds often appear dark when viewed through a telescope. Infrared light can penetrate molecular clouds, and indeed IR spectroscopy is a key method for identifying molecular species within these regions. However, infrared photons do not have sufficient energy to ionize neutral molecules. Instead, most ions within molecular clouds are formed through collisions with cosmic rays.

Evidence to date show that ice is mixed with dust in space. The examination of the samples returned by the Stardust mission showed that the dust and water ice agglomerates were mixed before cometesimals formed in the outer Solar system (Brownlee et al. (2006). Results of the Rosetta mission to comet 67P/Churyumov–Gerasimenko demonstrated that the core consists of a mixture of ices, iron sulfides, silicates and hydrocarbons (Fulle et al. 2017). Different models of comets, such as dirty snowball, icy glue, and fractal aggregates (A'Hearn 2011), mean that ice is mixed with dust in cometary nuclei. Asteroids are expected to retain ices in their interior. Dust grains in astrophysical environments form fractal aggregates characterized by a very high porosity that is suggested by the analysis of cometary dust particles, see Fulle et al. (2000), Fulle and Blum (2017), Harmon et al. (1997), Hörz et al. (2006), dust evolution models, see Kataoka et al. (2013), Ossenkopf (1993), Tazaki et al. (2016), and laboratory experiments on the dust particle aggregation by collisions, see Krause and Blum (2004), Wurm and Blum (1998), and on the gas-phase condensation of grains with their subsequent aggregation on a substrate, see Jäger et al. (2008), Sabri et al. (2014). It is reasonable to assume that ice fills the pores of grain aggregates.

The chemical evolution of galaxies reflects in the changes of chemical composition (relative abundances of chemical elements) of stars, interstellar gas, and dust. The elements heavier than helium are synthesized in stars and then ejected by a dying star. The next generation of stars forms from gas clouds that include heavy elements from the previous stellar generation. Therefore, stars in the present day galaxy are fossils that retain the information on the properties of stars from the past. It is, in principle, possible to find the chemical composition of a galaxy as a function of position and time by measuring the abundances of stars with different birthplaces and ages. This approach is valid, provided that their atmospheres represent the composition of the gas from which they were formed. Such studies may give valuable information about the chemical evolution of galaxies and even about the structure of matter in the very early phases of the Universe.

All dying stars return a fraction of all of their envelope mass to the interstellar medium (ISM) by stellar winds. These winds (for massive stars occurring before the final supernova explosions) carry newly processed metals and the unprocessed metals that were trapped inside the star at its formation and return them to the ISM.

Recently, Kobayashi et al. (2020) constructed Galactic Chemical Evolution (GCE) models for all stable elements from $^{12}$C to $^{238}$U from first principles by using theoretical nucleosynthesis yields and event rates of all chemical enrichment sources. This enables the prediction of the origin of elements as a function of time and environment. The basic equations of chemical evolution are described by Kobayashi et al. (2000). The code follows the time evolution of elemental and isotopic abundances in a system where the ISM is instantaneously well mixed. **Fig. 2** shows the theoretical age-metallicity (metallicity [Fe/H] is a logarithmic measure of the iron abundance relative to the sun) relationship for solar neighborhood stars. The time is in units of $10^9$ years.

Using the data and GCE model calculations for individual chemical elements to Fe concentration ratios as a function of metallicity [Fe/H], as presented in the work by Kobayashi et al. (2020), one can calculate concentration values ratios for both for life essential bulk elements (C: N: O: Na: Mg: P: S: Cl: K: Ca) and for essential trace elements (Cr: Mn: Fe: Co: Ni: Cu: Zn: Se: Mo: I: W) as a function of metallicity [Fe/H]. This should be done for different [Fe/H] values corresponding to different times, T-values.

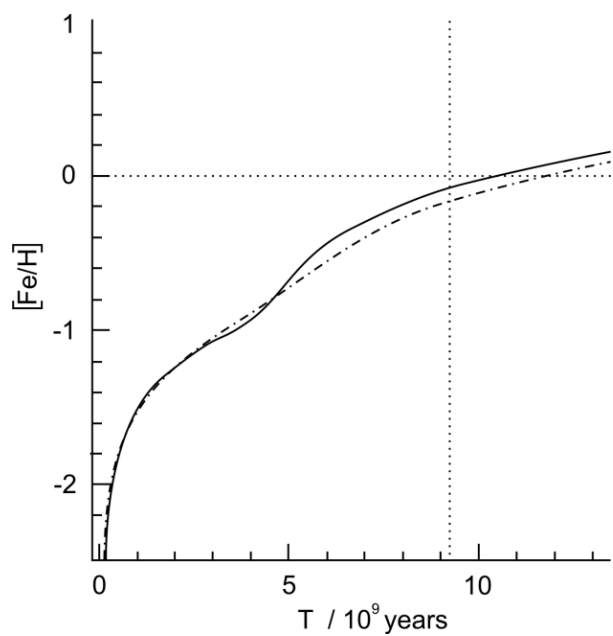

**Fig. 2**: Age-metalicity relation of solar neighborhood models by Kobayashi et al. (2020).

## 3. Chemistry of Life

Elemental requirements of present day life on the planet Earth are defined by the needs of only some of the elements for its existence. Life, as we know is (H-C-N-O) based, is relying on the number of bulk (Na-Mg-P-S-Cl-K-Ca) and trace elements (Li-B-F-Si-V-Cr-Mn-Fe-Co-Ni-Cu-Zn-As-Se-Mo-W). Trace elements are used by all organisms and provide proteins with unique coordination, catalytic, and electron transfer properties.

The main characteristic of living matter is the accumulation of chemical elements from the environment, often against concentration gradient. Concentration factors, ratios of chemical element concentrations in the living matter and environment, are important parameters for understanding the life processes.

Essential elements for life are:
(i) Four basic organic elements: H, C, N, O.
(ii) Essential bulk elements: Na, Mg, P, S, Cl, K, Ca.
(iii) Essential trace elements: Cr, Mn, Fe, Co, Ni, Cu, Zn, Se, Mo, I, W.
(iv) No confirmed functions in humans, but plants, animals, etc.: Li, B, F, Si, V, As.
(v) The rest of the chemical elements are environmental pollutants.

Assuming all organisms on the tree of life have a common ancestor i.e. tree is rooted in the LUCA, it remains to determine its nature. It is reasonable to assume that LUCA had simple metabolism processes, with few essential amino acids and perhaps few non-specific enzymes. In establishing an average elemental composition of life one needs to estimate the relative abundance of elements in LUCA. One such effort to derive an estimate for the elemental composition of LUCA has been discussed by Chopra et al. (2010) and Chopra and Lineweaver (2015). Their compilation samples eukaryotic, bacterial, and archaeal taxa across the extant tree of life taking into account elemental abundances and phylogenetic relationship between the taxa.

In evaluating the bulk elemental composition of LUCA, Chopra and Lineweaver (2015) used abundances for almost all of the biologically relevant elements, including the major elements (H, O, C, N, P, S) and minor elements (e.g., Na, K, Mg, Ca, Fe, Cu, Zn). In establishing an average elemental composition of life, they attempted to account for differences in structure between species and other phylogenetic taxa by weighting datasets such that the result represents the root of prokaryotic life (LUCA). Variations in composition between data sets that can be attributed to different growth stages or environmental factors were used as estimates of the uncertainty associated with the average abundances for each taxon. The idea of a LUCA of all cells, or the progenitor, is the most important for the study of early evolution and life's origin, yet information about how and where LUCA lived is still lacking. The physiology and possible habitat of LUCA have been recently discussed (Weiss et al. 2016).

There are several complexities in unraveling the functional capabilities of the LUCA. First, the open question of where the tree of life is rooted must be resolved as only then can it be determined whether high-temperature adaptation is a primary inherited trait. Second, even if the tree of life does not have a root in organisms adapted to living at high temperatures, this does not rule out a hot origin of life. Namely, considerable time and extinct organisms might exist between the origin of life and the root of the tree, as defined by extant organisms. All these considerations are done by assuming that the three existing domains of life define the LUCA's capabilities and characteristics. However, as pointed out by Cockell (2015), we cannot be sure that there were not domains that went extinct early in the history of life and took with them crucial information. The best rating for life-essential elements of LUCA, as estimated by Chopra and Lineweaver (2015), is shown in **Fig. 3**.

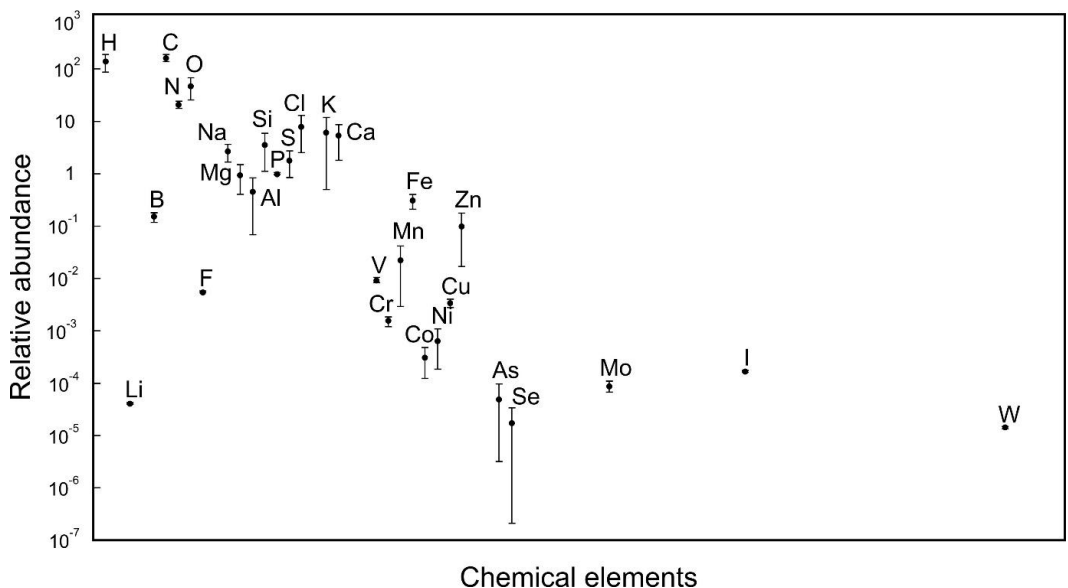

**Fig. 3**: The bulk elemental composition of LUCA as estimated by Chopra and Lineweaver (2015) from abundances in extant taxa. Only for life-essential elements are shown.

Quantitative estimates of the gene complement of LUCA may be masked by ancient horizontal gene transfer events and polyphyletic gene losses. Biases in genome databases and methodological artefacts also represent drawbacks. In spite, most reports agree that LUCA resembled extant prokaryotes. A number of highly conserved genes are sequences involved in the synthesis, degradation, and binding of RNA, including transcription and translation. Although the gene complement of LUCA includes sequences that may have originated in different epochs, the outstanding conservation of RNA-related sequences supports the hypothesis that the LUCA was an evolutionary result of the so-called RNA-protein world. The available evidence suggests that the LUCA was not a hyperthermophile. However, according to the current conventional understanding, it is not possible to assess LUCA's ecological niche or its mode of energy acquisition and carbon sources (Becerra 2007).

## 4. IMCs - Environment Where Llife Might Originated

Interstellar dust is coupled to gas clouds and, as such, carried around the Milky Way. These clouds come in a wide variety of shapes, sizes, densities, and temperatures. They can, however, be qualitatively classified into two basic categories: interstellar diffuse clouds and interstellar molecular clouds (IMCs). The diffuse clouds are not distinguishable and are limited to a density less than about 300 hydrogen atoms per $cm^3$ with a temperature of 50–100 K. The molecular clouds can have temperatures of 20 K (even 10 K) and density above 300 per $cm^3$ including the densities at which clouds collapse to form stars. Diffuse clouds contain hydrogen in the atomic form, whereas molecular hydrogen is a dominant component of molecular clouds. Molecular clouds can be very dark, as illustrated by the Horsehead nebula, and also by isolated blank regions in the sky widely called Bok globules. These latter are probably concentrations of dust and gas, which are collapsing to form stars (Greenberg 2002).

The energetic components of the environment affecting the dust (other than that associated with phenomena like star formation and even more explosive supernovae) are the ultraviolet radiation from stars and the cosmic ray particles. One of the significant effects of the ultraviolet is that it heats the dust. The dust temperature in diffuse clouds could be about 15 K as a result of a dynamic balance between absorption and emission of radiation. However, in molecular clouds where it is shielded, the temperature may be as low as 5–10 K (Greenberg 1971; Greenberg and Li 1996).

It appears that solid state chemistry is more relevant in the prebiotic context than thus far anticipated, see for example Becker et al. (2018), Bolm et al. (2018), Forszthe et al. (2015), Lamour et al. (2019), Lherminier et al. (2019). It has been shown (Stolar et al. 2020) that nucleobases methylated at the glycosidic nitrogen atom achieve DNA-specific self-assembly upon heating in the solid state. This pairing of nucleotides in the DNA via specific hydrogen-bonding interactions constitutes the most famous example of supramolecular recognition. Oba et al. (2019) reported the simultaneous detection of all three pyrimidine (cytosine, uracil, and thymine) and three purine nucleobases (adenine, xanthine, and hypoxanthine) in interstellar ice analogs composed of simple molecules including $H_2O$, CO, $NH_3$, and $CH_3OH$ after exposure to ultraviolet photons followed by thermal processes, that is, in conditions that simulate the chemical processes accompanying star formation from molecular clouds. Their results strongly suggest that the evolution from molecular clouds to stars and planets provides a suitable environment for nucleobase synthesis in space.

The recent findings of Obhodas et al (2020) demonstrated that magnetic field intensity of 100 nT enhances the growth of *Bacillus subtilis* an order of magnitude above controls that were growing in the Earth's magnetic field (~ 48 µT). This might suggest that microorganisms favor extremely low magnetic fields. The Earth's magnetic field of the approximately same intensity as presently (between 33 µT at the magnetic equator, and 67 µT at the magnetic poles), formed 4.2 x $10^9$ years ago, almost immediately after the lunar-forming giant impact (Tarduno *et al.*, 2020). It has been also shown that microorganisms in a very similar way favor microgravity (Horneck et al. 2010). These findings of microorganisms favoring the low magnetic and gravitational fields, evident as 10 times increase in their multiplication rate, strengthen the 2. hypothesis that life might originate in IMCs, where maximum magnetic field intensities are ~ 100 nT (Cruttcher *et al.,* 2012) and strong gravitational fields are not yet formed. The 2. hypothesis assumes that microorganisms would be protected from detrimental cosmic radiation by growing on the surface or within the dust particles.

## 5. Chirality phenomenon

Together with a problem of trace element essentiality, we should consider a closely related problem of chirality. A living organism, simply called life, is an organized system of molecules having specific handedness, called chirality. Chiral molecules are designated D (dextrorotatory) or L (levorotatory) according to the right or left direction, respectively, in which the crystalline forms rotate polarized light. Biological polymers (e.g. nucleic acids and proteins) use almost exclusively D-sugars and L-amino acids. The exception is glycine which is the only achiral proteinogenic amino acid. Laboratory synthesis made from optically inactive starting materials yields racemic (1:1) mixtures of L- and D- isomers. 19 of the 20 amino acids used to synthesize proteins can exist as L- or D- enantiomorphs. The glycine is the only exception as it has two (indistinguishable) hydrogen atoms attached to its alpha (central) carbon. The function of a protein is determined by its shape. A protein with a D amino acid instead of L will have its R group sticking out in the wrong direction (for details see for example LibreTexts 2020). Figs. 4 and 5 show examples of D and L amino acids and sugars, respectively.

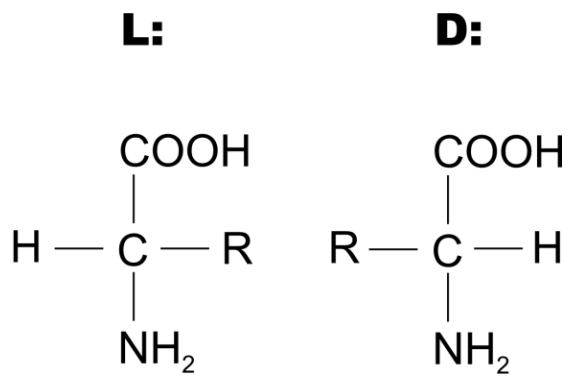

Fig. 4: Two enantiomers (left-handed - L and right-handed - D) of generic chiral amino acid.

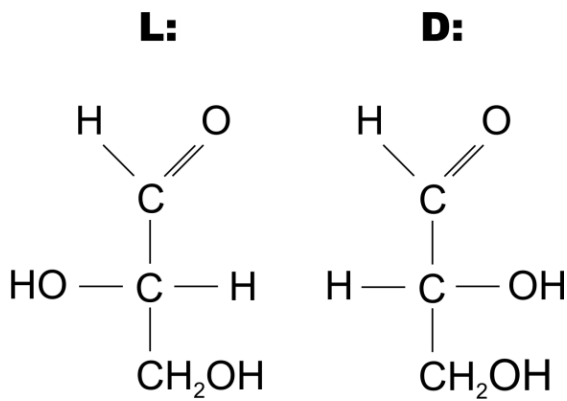

Fig. 5: Glyceraldehyde has one chiral center and therefore exists as two different enantiomers with opposite optical rotation.

The process of demarcation of the origins of the chemistry of life starts with evaluating the possibility of potential mechanisms for the assembly of molecules into polymers able of self-replication and transmittance of genetic information. At some point along this pathway, the single chirality of sugars and amino acids emerges as their hallmark in biological molecules. Researchers have developed abstract mathematical theses for the origin of homochirality from a presumably racemic collection of prebiotic molecules. Before the end of the 20$^{th}$ century, experimental findings supported several basic features of these scenarios, although these studies considered chemical systems without direct prebiotic relevance. Currently, researchers are examining plausible conditions that couple chemical and physical processes leading to a one-chirality of sugars and amino acids followed by chemical reactions enhancing molecular complexity. While these studies have been conducted in the frame of the "RNA World" scenario, the experimental findings support more a "metabolism first" scenario of the origin of life. Hein and Blackmond, (2012) incorporated both chemical and physical phenomena that allow for the amplification of a small initial imbalance of either sugar by amino acids or amino acid by sugars. They suggested that an enantio-enriched chiral pool of one type of molecule could lead to a similarly enantio-enriched pool of the other.

Considering all that has been said above, one might ask an obvious question: Could there have been some asymmetric factors in the environment that influenced the choice when life began? There have been several suggestions, including 1) polarized light, 2) optically active quartz, 3) natural radioactivity. Irradiation of racemic system by circularly polarized radiation mechanisms can lead to small enantiomeric excess (L-enantiomeric enrichments of 0.2–2.5%) (Modica et al. 2014). The last scheme assumes the origin of life in a radioactive environment. One possibility is e-radioactivity resulting in longitudinally polarized β-rays, which produce circularly polarized bremsstrahlung. This radiation was considered to have preferentially destroyed D-isomers. Some recent experiments have not supported this assumption (Bonner 2020). We propose a new mechanism that might result in chirality excess, assuming that amino acids were synthesized on dust particles in interstellar space. This is the bombardment by cosmic rays of high energy polarized protons which preferentially destroy one isomer because of significant asymmetry in proton (in cosmic rays)-proton (in amino acid) scattering. This will be explained in more detail in the next chapters.

Interstellar dust appears to play a critical role in the formation of interstellar molecules. Molecules may be formed on or in grain surfaces. Many molecules have been observed in interstellar space. Johnson (1972) has long ago reported the observation of interstellar porphyrins (molecule $MgC_{46}H_{30}N_6$). The existence of interstellar molecules suggests:

1) Such molecules support or are the metabolic products of an interstellar biota.

2) Such molecules, participating in planetary condensation from the interstellar medium, can make a significant contribution to the origin of terrestrial life.

The elongated/irregular dust particles in the dust clouds are aligned by the magnetic fields. Such dust particle alignment has been observed for solar magnetic field and it has assumed to be the reason why the light scattered by cometary dust becomes circularly polarized (CP) (Kolokolova et al. 2015). Actually, the most popular explanation of the CP formation is the scattering of light on aligned/irregular dust particles, or on the particles that contain homochiral molecules. The observed chirality might be a consequence of the preferential destruction of one enantiomer on the surface of aligned dust particles.

As early as 1984, Joyce et al. suggested that for life to emerge, something first had to crack the symmetry between left-handed and right-handed molecules, an event biochemist call "breaking the mirror". Since then, the search for the origin of life's handedness in the prebiotic worlds has mainly been focused on physics and chemistry, not biology. Joyce et al. (1984) showed that the polymerization of activated mononucleotides proceeded readily in a chiral system. However, it is severely inhibited by the presence of the opposing enantiomer. This finding poses a challenge for the spontaneous emergence of RNA-based life. It has been suggested that (i) RNA was preceded by some other genetic polymer that is not subject to chiral inhibition or (ii) chiral symmetry was broken through chemical processes before the origin of RNA-based life.

Once an RNA enzyme arose that could catalyze the polymerization of RNA, it would be possible to distinguish among the two enantiomers, enabling the process of RNA replication and the start of RNA-based evolution. It is assumed that the earliest RNA polymerase and its substrates have been of the same handedness, but this might not be the case. Replication of D- and L-RNA molecules may have emerged together, assuming the ability of structured RNAs of one handedness to catalyze the templated polymerization of activated mononucleotides of the opposite handedness, Sczepanski and Joyce (2014).

Recently Yin et al. (2015) found that any molecule, if large enough (several nanometers) and with an electrical charge, will seek their type with which to form a large assembly. The authors have shown that homochirality, or how molecules select other like molecules to create larger assemblies, may not be as mysterious as previously imagined. The remaining problem is understanding how homochirality occurred at the onset of life. If chirality emerged sometime after the origins of life, the question remains: Why did right-handed RNA win? Left- and right-handed molecules have chemically identical properties, so there is no apparent reason for one to triumph. Yin et al. (2015) have shown that chiral macro-anions demonstrate chirality recognition behavior by forming a homogeneous blackberry structure via long-range electrostatic interactions between the individual enantiomers in their racemic solutions. Adding chiral co-anions suppresses the self-assembly of one enantiomer while maintaining the assembly of the other one. This process leads to a natural chirality selection and chirality amplification process, indicating that some environmental preferences can lead to a complete chirality selection. The fact that the relatively simple inorganic macro-ions exhibit chirality recognition and selection during their assembly process indicates that the related features of bio-macromolecules might be due to their macro-ionic nature via long-range electrostatic interactions.

We should here discuss the findings related to meteorite Asuka 12236. Its origin fits toward the very beginning of the solar system. In fact, some scientists think that the tiny pieces of the meteorite predate the solar system. Several lines of evidence suggest that Asuka 12236's original chemical makeup is the best preserved in a category of carbon-rich meteorites known as CM chondrites. These are very interesting rocks for the study of the origin of life since many of them contain a highly complex mixture of organic compounds associated with living matter. The interior of Asuka 12236 is so well preserved because the rock was exposed to very little water or heat, both when it was still a part of an asteroid and later when it sat in Antarctica waiting to be discovered. There are two facts:
1. there are no clay minerals inside the material, and
2. iron metal in it has not rusted (no exposure to oxygen in the water).

The rock also contains an abundance of silicate grains with unusual chemical compositions that indicate they formed in ancient stars that died before the Sun began to form. The liquid water and heat are needed to produce a variety of amino acids, however, too much of both can destroy them all. It turns out that the inside of an asteroid has perfect conditions for this process. The water would have been produced inside the asteroid that Asuka 12236 came from, as heat from the radioactive decay of certain chemical elements melted the ice that condensed with a rock when the asteroid first formed. Given that Asuka 12236 is so well preserved it could have come from a cooler outer layer of the asteroid where it would have come in contact with little heat, and thus, water. Glavin et al. (2020) found a huge excess of left-handed molecules than right-handed ones in some protein-building amino acids in Asuka 12236 (left enantiomeric excess $L_{ee} \approx 34\text{-}64\%$). Since there are no evidence for biological amino acid contamination, it is possible that some of the L-enantiomeric excesses of the protein amino acids, aspartic and glutamic acids, and serine, are indigenous to the meteorite. None of the possible physical processes (e.g. polarized light, optical activity of quartz, natural radioactivity, or cosmic rays) can explain such a large excess of L amino acids found in the Asuka 12236, and amplification mechanisms must be sought (Glavin et al. 2020). Should the amino acids found on the Asuka 12236 meteorite be biogenic and have not been exposed to terrestrial contamination, they would have existed long before the appearance of life on Earth. It is noteworthy to mention that many carbonaceous meteorites show a slight increase of L amino acids (Pizzarello 2006) or D sugars (Cooper and Rios 2016)

## 6. Iterstellar space: cosmic rays and magnetic fields

Primary cosmic rays are stable charged particles that have been accelerated to enormous energies by astrophysical sources somewhere in our Universe. They must be stable in order to survive the long trips through interstellar (or intergalactic) space. They are charged because the accelerating mechanism is probably electromagnetic and because their charge is what interacts with matter and produces the effects that we can observe. They have a range of energies, $10^9$ eV (1GeV) up to $10^{20}$ eV ($10^8$ TeV), for comparison, the latest accelerators here on earth can achieve only about 7 TeV. The most common primary cosmic ray particle is the proton or hydrogen nucleus. 95% of all cosmic rays are protons, 4% are helium nuclei, and the 1% balance is made up of nuclei from elements up to iron.

Hillas (1967) suggested that the output of cosmic ray sources must have been very much greater in the past than at present. Attention should be drawn to energy losses suffered by the most energetic cosmic ray protons in passing through the flux of microwaves, corresponding to blackbody radiation of 3 K, which probably permeates space. Energy loss is due to pairing production by microwave photons colliding with high energy protons modifying the spectrum of cosmic rays. One has to take into account the energy loss due to the expansion of the Universe (the "red-shift" effect).

Protons in the cosmic rays get scattered of different targets on their path through the Universe and along with the magnet fields force, including individual atoms in the molecules in molecular clouds and on the dust particle surfaces. Among other processes, the one process of interest is elastic proton-proton scattering as shown in Fig. 6. Either proton (beam or target) can be polarized.

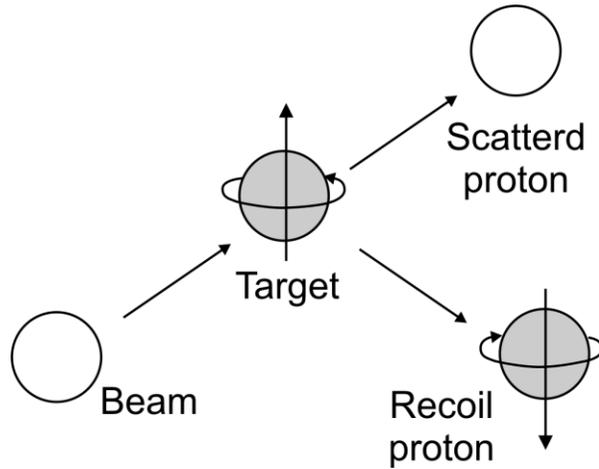

Fig. 6: The elastic scattering process. $A_N$ arises from the interference between a spin-flip (coulomb) and spin-non flip (Nuclear) amplitude.

Analyzing power $A_N(t)$ for p-p elastic scattering in the energy range of energetic cosmic ray protons has been studied by several groups, see Akchurin et al. (1993), Bravar et al. (2005), Okada et al. (2008). Some of this work is summarized by Bazilevsky et al. (2011) and presented in Fig. 7. The analyzing power ($A_N$) can be extracted from the asymmetry between the number of scatterings on the left versus on the light, corrected for the left and right detector acceptances, or from the asymmetry between the number of scattering (e.g. on the left) for target polarization up and target polarization down, corrected for the integrated luminosities for the corresponding target spin states. These two approaches can be combined in so called "sqrt" formula, which cancels contributions from different left-right detector acceptances and different luminosities in the measurements with up and down target polarization states to the asymmetry:

$$A_N = \frac{1}{P_T} \frac{\sqrt{N_L^\uparrow \times N_R^\downarrow} - \sqrt{N_R^\uparrow \times N_L^\downarrow}}{\sqrt{N_L^\uparrow \times N_R^\downarrow} + \sqrt{N_R^\uparrow \times N_L^\downarrow}},$$

where $N_{L(R)}^{\uparrow(\downarrow)}$ is the number of recoil protons selected from p-p elastic scattering events detected on the left (right) side of the beam. $P_T$ is target polarization and the arrows give the direction of the target polarization. The $A_N$ measurements were performed for the recoil proton kinetic energy range $T_R$ = 1 - 4 MeV corresponding to a momentum transfer $0.002 < -t < 0.008$ $(10^9$ eV/c$)^2$; $-t = 2m_p T_R$, where $m_p$ is the proton mass.

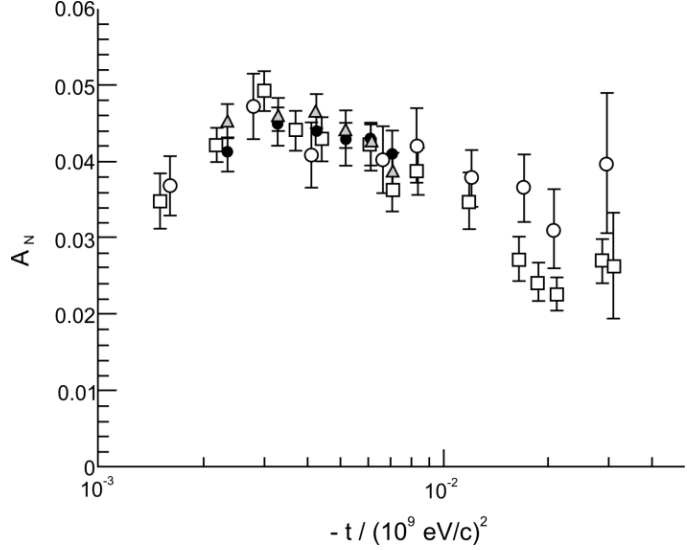

Fig. 7: Analyzing power ($A_N$) for elastic p-p scattering: 24 GeV (open circles) and 100 GeV (open squares, $s^{1/2}$ = 6.8 and 13.7 GeV), 31 GeV (closed circles) and 250 GeV (shaded triangles, $s^{1/2}$ = 7.7 and 21.7 GeV) all data are from different Relativistic Heavy Ion Collider (RIHC) runs, The uncertainties shown are the quadratic sum of the statistical and systematic uncertainties, after Bazilevsky et al. 2011.

The values of the mean polarizations determined with the EDDA polarimeter averaged over all data at the beam energy $T_p$ where the p-p analyzing power was measured in ANKE, see Table 1. Though the shown statistical errors are small, there are 3% systematic uncertainties. The normalization factors N are those obtained in a partial wave fit to the current STT data, after Bagdasarian et al. 2014.

Table 1. p-p analyzing power for different bombarding energies.

| $T_p$ (MeV) | 796 | 1600 | 1800 | 1965 | 2157 | 2368 |
|---|---|---|---|---|---|---|
| p | 0.554 ±0.008 | 0.504 ±0.003 | −0.508 ±0.011 | −0.429 ±0.008 | −0.501 ±0.010 | 0.435 ±0.015 |
| N | 1.00 | 1.00 | 0.99 | 1.09 | 1.01 | 0.93 |

## 7. Preferential destruction of enantiomers

For DNA today, radiation increases the frequency of gene mutations; this has been known since the pioneering work of Muller (1927) that showed that the mutation rate is proportional to the radiation dose, much of it attributable to ionization by cosmic rays. Ionization by spin-polarized radiation could be enantio-selective (Zel'dovich et al. 1977). Therefore, one can argue that the mutation rate of L- and D- organisms would be different, see Globus and Blandford (2020). As there could be $10^9$ or even $10^{12}$ of generation of the earliest and simplest life forms, a small difference in the mutation rate could easily sustain one of the two early, chiral choices.

On other hand, laboratory experiments have demonstrated that it is possible to induce an enantiomeric excess of amino acids by irradiation of interstellar ice analogs with UV CPL (de Marcellus et al. 2011). However, this raises two problems. First, circular dichroism is also wavelength-, pH-, and molecule-specific (D'hendecourt et al. 2019). It is hard to see how one sense of circular polarization can enforce a consistent chiral bias, given the large range of environments in which the molecules are found. Second, it is often supposed that astronomical sources supply the polarization. If we seek a universal, chiral light source that consistently emits one polarization over another, then we are again drawn to the weak interaction in order to account for a universal asymmetry. One option is to invoke spin-polarized particles, which can radiate one sense of circular polarization through Čerenkov radiation or bremsstrahlung and can preferentially photolyze chiral molecules of one handedness. Another option is to invoke supernova neutrinos (Boyd et al. 2018). However, the small chiral bias is unlikely to lead to a homochiral state and some prebiotic amplification mechanism is still required. This suggests enantio selective bias in the evolution of the two living systems. Although the two forms of a chiral molecule have identical physical and chemical properties, the interaction with other chiral molecules may be different. Chiral molecules in living organisms exist almost exclusively as single enantiomers, a property that has a critical role in molecular recognition and replication processes. Therefore, it is a prerequisite for the origin of life. Left-handed and right-handed molecules of a compound will be formed in equal amounts (a racemic mixture) when synthesized in the laboratory in the absence of some directing template.

The existence of the single chirality of biological molecules, almost exclusively left-handed amino acids and right-handed sugars, presents us with two questions (Blackmond 2010):

(i) What served as the original template for the biased production of one enantiomer over the other in the chemically austere and racemic environment of the prebiotic world?

(ii) How was this sustained and propagated to result in the biological world of single chirality that surrounds us?

A review by Blackmond (2010) focuses primarily on the second question: the plausible mechanisms for the evolution of molecular chirality as exemplified by the D-sugars and L-amino acids found in living organisms today. "Symmetry breaking" is the term used for describing the occurrence of a difference between left and right enantiomeric molecules. This imbalance is measured in terms of the enantiomeric excess, or *ee*, where *ee = (R-L)/(R+L)*. *R* and *L* stand for concentrations of the right and left-hand molecules, respectively. Speculations for how an imbalance might have come about could be grouped as either terrestrial or extraterrestrial, and further subdivided into either random or deterministic. Evidence of L enantiomeric excess in amino acids found in chondritic meteor deposits (Pizzarello 2006) allows the hypothesis that the initial imbalance is older than our world.

Assuming that amino acids, and some sugars, could be synthesized on dust particles in interstellar space, the observed optical activity may be a result of cosmic ray bombardment. Two processes could be involved:

(i) The first process might arise from asymmetric dust grains aligned with the magnetic field. Interstellar medium (ISM) polarization by aligned dust grains has been probed from the diffuse medium in the UV to heavily obscured sightlines in the near-IR and into dense clouds and heavily embedded sources using far-IR emission (for details see Andersson et al. 2015). It should be kept in mind, that not all environments produce the same amount of polarization. Although, the dust mainly consists of silicates, amorphous carbon, and small graphite particles in various simple shapes, in the dark cold parts of molecular clouds grain aggregates and mantles of volatile ice might form. Only a limited number of relatively large grain sizes (~0.01-1µm) contribute to the polarization. This situation corresponds to a polarized target in the p-p elastic scattering process.

(ii) Alternatively, consider polarized beam interaction could be considered. High energy polarized protons in cosmic rays may be able to preferentially destroy one isomer on the dust particle surface because of significant asymmetry in proton (in cosmic rays) - proton (chirality center in amino acid or sugar) scattering. See Fig. 8. for the schematic presentation of the process leaeding to preferential destruction of D-amino acids and L-sugars.

Single spin asymmetry in polarized p-p elastic scattering has been experimentally studied at high energies, see, for example, Adamczyk et al. (2013) for experiments done at Relativistic Heavy Ion Collider, RHIC. RHIC is one of only two operating heavy-ion colliders, and the only spin-polarized proton collider ever built. It is located at Brookhaven National Laboratory (BNL) in Upton, New York. Polarization produced in p-p scattering has been observed even at low energies, 30 and 50 MeV (Batty et al. 1963).

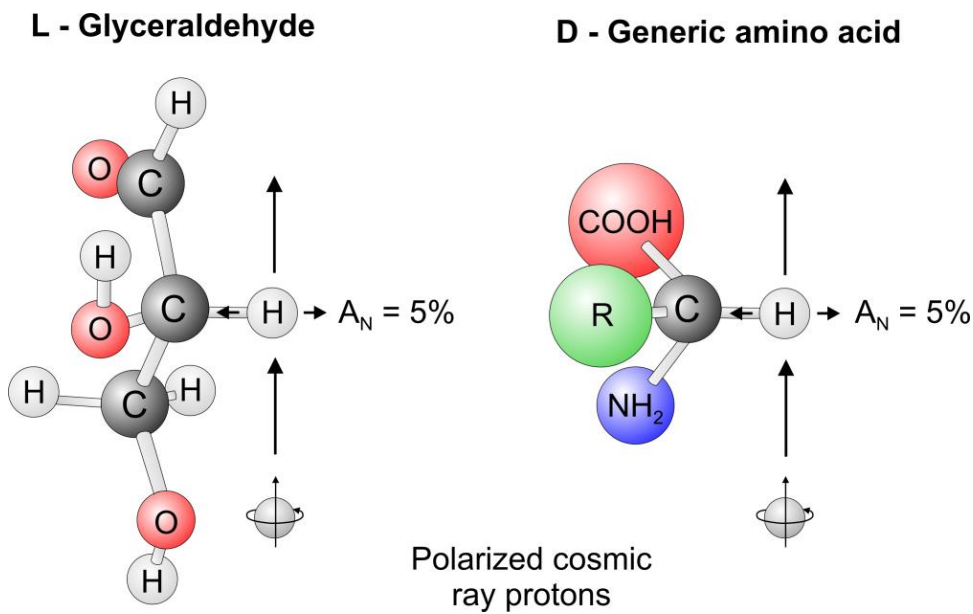

Fig. 8: Schematic presentation of the process of preferential destruction of D-amino acids and L-sugars.

If a sufficient enantiomeric excess existed in life's primordial molecular reservoir, it would almost certainly push life toward extreme bias of actual living beings (McGuire and Carroll 2016). Meteoritic samples suggest that such an enantiomeric excess has been generated before Earth formation. The authors McGuire and Carroll (2016) present a case of a meteorite that struck just outside of the city of Murchison, Victoria, Australia, on 28 September 1969. Its weight was ~100 kg, and it belongs to the class of Carbonaceous chondrite and rich in organic compounds, including amino acids, many of them having an excess of left-handed enantiomers > 10%. It has been determined that meteorites, like the one that hit Murchison, date back to about the time of our solar system's formation. Also, the source of their molecular material can be traced to the cloud of gas and dust from which our Solar system formed. There is a possibility of a reasonable mechanism existence for generating an enantiomeric excess in that primordial cloud. In that case, the origins of life's enantiomeric bias could be linked to processes that occurred billions of years ago, before the Solar system existed (McGuire and Carroll, 2016), see also Engel and Macko (1997) and McGuire et al. (2016).

Observation of CP formation in light passage through dust cloud would be a direct evidence of homochiral molecules existence within dust cloud (Gronstal 2014). Since the life on planet Earth is chiral, it is reasonable to assume that dust that formed our Solar system contained already this information. At this point we can put forward our **4.** *hypothesis: Chirality is a sine qua non condition for the emergence of life.*

## 7. Conclusions

We have presented numerous examples of observations, experiments, and theoretical considerations which we have used to synthesize four hypotheses concerning the origin of life. The Universe was born only with plenty of hydrogen isotopes, some helium, and lithium, while all other elements were formed as the Universe aged through star formation processes, their lives, and deaths, which resulted in dust clouds as a birthplace of a new generation of stars. During this process of the Universe aging the chemical element abundance curve has been changing in such a manner that at the time $T_{origin}$ coincided with the abundance curve of living matter.

Our two hypotheses, #1 and #3, define this time, $T_{origin}$, as a time when conditions were right for life to originate in the primitive form and that happened only once in the history of the Universe. The other two hypotheses, #2 and #4, define the dust particles in the molecular clouds and their environment as a place where this has happened. Exposure to cosmic rays and magnetic fields as well as nuclear physics laws made life chiral.

**References:**

Adamczyk, L. et al. 2013. Single spin asymetry $A_N$ in polarized proton-proton elastic scattering at √s=200 GeV. Physics Letters B 719: 62-69.

Adams, W. S. 1949. Observations of Interstellar H and k, Molecular Lines, and Radial Velocities in the Spectra of 300 O and B Stars. Astrophysical Journal, vol. 109, p.354-379

A'Hearn, M. F. 2011. Comets as building blocks. *Annu. Rev. Astron. Astrophys.* 49: 281–299.

Akchurin, N.,et al.,1993. Analyzing power measurement of pp elastic scattering in the Coulomb-nuclear interference region with the 200-GeV/c polarized-proton beam at Fermilab. Phys Rev. D 48: 3026.

Andersson, B.-G., Lazarian, A. and Vaillancourt J. E. 2015. Interstellar Dust Grain Alignment. Annual review of Astronomy and Astrophysics 53: 501-540.

Bagdasarian, Y. et al. 2014. measurement of the analzsing power in proton/proton elastic scattering at small angles. Phzsics Letters B 739> 152/156.

Batty, C. J., Gilmore, R. S., Stafford, G. H. 1963. Measurement of the polarization in proton-proton scattering at 30 and 50 MeV. Nuclear Physics 45:481-491.

Bazilevsky, A. et al. 2011. measurements of the energy dependence of the analyzing power in pp elastic scattering in the CNI region. Journal of Physics: Conference Series 295: 012096.

Becerra, A., Delaye, L., Islas, S. and Lazcano, A. 2007. The Very Early Stages of Biological Evolution and the Nature of the Last Common Ancestor of the Three Major Cell Domains. Annual Review of Ecology, Evolution, and Systematics 38: 361-379.

Becker, S., Schneider, C., Okamura, H., Crisp, A., Amatov, T., Dejmek, M., Carell, T. 2018. Wet-dry cycles enable the parallel origin of canonical and non-canonical nucleosides by continuous synthesis. Nature Communition 9(1): 163.

Blackmond, D. G. 2010. The Origin of Biological Homochirality. Cold Spring Harbour Perspectives in Biology (Eds.Deamer, D and Szostak, J. W.) 2:a002147.

Bolm, C., Mocci, R., Schumacher, C., Turberg, M., Puccetti, F., and Hernandez, J.G. 2018. Mechanochemical Activation of Iron Cyano Complexes: A Prebiotic Impact Scenario for the Synthesis of α-Amino Acid Derivatives.  Angew. Chem., Int. Ed. 57: 2423-2426.

Bonner, W.A. 2020. Parity violation and the evolution of biomolecular homochirality. Chirality 12(3):114-126.

Boyd, R. N., Famiano, M., Onaka, T. and Kajino, T. 2018. Sites That Can Produce Left-Handed Amino Acids in the Supernova Neutrino Amino Acid Processing Model. The Astrophysical Journal 856(1): article id. 26, 5 pp.

Bravar, A. et al. 2005. Spin Dependence in Elastic Scattering in the CNI Region. Brookhaven National Laboratory. Report BNL-75128-2005-CP.

Brownlee, D. et al. 2006. Comet 81P/Wild 2 under a microscope. *Science* 314: 1711–1716.

Cassé, M., Lehoucq, R. and Vangloni-Flam, E. 1995. Production and evolution of light elements in active star-forming regions. Nature 373: 318 – 319.

Cazaux, S., Minissale, M., Dukieu, F., Hocuk, S. 2016. Dust as interstellar catalyst II. How chemical desorption impacts the gas. A&A 585, A55.

Chopra, A., Lineweaver, C. H., Brocks, J. J. and Ireland, T. R. 2010. Palaeoecophylostoichiometrics: Searching for the Elemental Composition of the Last Universal Common Ancestor. Australian Space Science Conference Series: 9th Conference Proceedings, Sidney, 28-30 September 2009 (DV). National Space Society of Australia Ltd, ISBN 13: 978-0-9775740. (www.tinyurl.com/ACetal10).

Chopra, A. and Lineweaver, C. H. 2015. An Estimate of the Elemental Composition of LUCA. Astrobiology Science Conference 2015. Chicago, Illinois, June 15-19, 2015. Paper 7328.

COBE, Cosmic Background Explorer. 2015. http://lambda.gsfc.nasa.gov/product/cobe/.

Cockell, C. 2015. Astrobiology – Understanding Life in the Universe. John Wiley & Sons Ltd. The UK. A companion website: www.wiley.com/go/cockell/astrology.

Cooper, G. and Rios, A. C. 2016. Enantiomer excesses of rare and common sugar derivatives in carbonaceous meteorites. PNAS, www.pnas.org/cgi/doi/10.1073/pnas.1603030113. (10 pp).

Crutcher RM. 2012. Magnetic Fields in Molecular Clouds. Annual Review of Astronomy and Astrophysics. 50(1):29–63. doi:10.1146/annurev-astro-081811-125514

de Marcellus, P., Meinert, C., Michel Nuevo, M. et al. 2011. Non-racemic amino acid production by ultraviolet irradiation of achiral interstellar ice analogs with circularly polarized light. The Astrophysical Journal Letters, 727: L27 (6pp).

D'Hendecourt L., Modica P., Meinert C., Nahon L., Meierhenrich U. 2019. Interstellar ices: a possible scenario for symmetry breaking of extraterrestrial chiral organic molecules of prebiotic interest. arXiv:1902.04575v1 [astro-ph.EP].

Draine, E. T. 2003. Interstellar dust grains. Annu. Rev. Astron. Astrophys. 41: 241-289.

Ehrenfreund, P. and Charnley, S. B. 2000. Organic Molecules in the Interstellar Medium, Comets, and Meteorites: A Voyage from Dark Clouds to the Early Earth. Annual Review of Astronomy and Astrophysics 38(1): 427-483.

Engel, M. H. and Macko, S. A. 1997. Isotopic evidence for extrater- restrial non-racemic amino acids in the Murchison meteorite. Nature 389: 265-268.

Forsythe, J. G., Yu, S.-S., Marnajanov, I., Grover, M. A., Krishnamurthy, R., Pernandez, E. M., and Hud, N. V. 2015. Ester-Mediated Amide Bond Formation Driven by Wet–Dry Cycles: A Possible Path to Polypeptides on the Prebiotic Earth. Angew. Chem., Int. Ed. 54: 9871-9875.

Fulle, M. et al. 2017. The dust-to-ices ratio in comets and Kuiper belt objects. *Mon. Not. R. Astron. Soc.* 469: S45–S49.

Fulle, M., Levasseur-Regourd, A. C., McBride, N. & Hadamcik, E. 2000. In situ dust measurements from within the coma of 1P/Halley: first-order approximation with a dust dynamical model. *Astron. J.* 119: 1968–1977.

Fulle, M. & Blum, J. Fractal dust constrains the collisional history of comets. *Mon. Not. R. Astron. Soc.* 469: S39–S44.

Glavin, D. P. et al. 2020. Abundant extraterrestrial amino acids in the primitive CM carbonaceous chondrite Asuka 12236. Meteoritics & Planetary Science, 20 August 2020 (28 pp), doi: 10.1111/maps.13560.

Globus, N. and Blandford, R. D. 2020. The Chiral Puzzle of Life. The Astronomical Journal Letters 895: L11 (14pp).

Greenberg, J.M. 1971. Interstellar Grain Temperatures: Effects of Grain Materials and Radiation Fields. Astron. Astrophys. 12: 240-249.

Greenberg, J. M. 2002. Cosmic dust and our origins. Surface Science 500: 793–822.

Greenberg, J. M. and Li, A. 1997. Grain temperatures and emission in diffuse interstellar clouds. Diffuse Infrared radiation and the IRTS, ASP Conference Series, Vol. 124, 1997. Okuda, H Mataumoto, T. and Roelling, T. L. Eds.

Gronstal, A. L. 2014. Light scattering on dust holds clues to habitability. Astrobiology Magazine, Sep. 25. 2014. https://www.astrobio.net/alien-life/light-scattering-dust-holds-clues-habitability/.

Güsten, R., Wiesemeyer, H., Neufeld, D. et al. 2019. Astrophysical detection of the helium hydride ion HeH+. Nature 568: 357–359.

Harmon, J. K. et al. 1997. Radar detection of the nucleus and coma of comet Hyakutake (C/1996 B2). *Science* 278: 1921–1924.

Hein, J. E. and Blackmond, D. G. 2012. On the Origin of Single Chirality of Amino Acids and Sugars in Biogenesis. Acc. Chem. Res. 45(12): 2045–2054.

Hillas, A. M. 1967. Cosmic rays in an evolving universe. Presented at 10th Inter. Conf. on Cosmic Rays, Calgary, June 19-30, 1967. Papr OG-71.

Hörz, F. et al. 2006. Impact features on Stardust: implications for comet 81P/Wild 2 dust. *Science* 314, 1716–1719.

Jäger, C., Mutschke, H., Henning, T. and Huisken, F. 2008. Spectral properties of gas-phase condensed fullerene-like carbon nanoparticles from far-ultraviolet to infrared wavelengths. *Astrophys. J.* 689: 249–259.

Johnson, F.M. 1972. Interstellar Matter II: Diffuse Interstellar Lines and Porphyrins. Ann. New York Acad. Sci. 187: 186-206.

Joyce, G., Visser, G., van Boeckel, C. et al. 1984. Chiral selection in poly(C)-directed synthesis of oligo(G). Nature 310: 602–604. https://doi.org/10.1038/310602a0.

Kataoka, A., Tanaka, H., Okuzumi, S. & Wada, K. 2013. Fluffy dust forms icy planetesimals by static compression. *Astron. Astrophys.* 557: L4.

Kobayashi, C., Tsujimoto, T. and Nomoto, K. 2000. The History of the Cosmic Supernova Rate Derived from the Evolution of the Host Galaxies. The Astrophysical Journal 539: 26-38.

Kobayashi, C., Karakas, A. I. and Lugaro, M. 2020. The Origin of Elements from Carbon to Uranium. The Astrophysical Journal 900: 179 (33pp).

Kolokolova, L., Koenders, C., Rosenbush, V., Kiselev, N., Ivanova, A., Afanasiev, V. 2015. Dust Particle Alignment in the Solar Magnetic Field: a possible Cause of the Cometary Citcular Polarization.American geophysical Union, Fall Meeting 2015, abstract id P41D-2089.

Krause, M. and Blum, J. 2004. Growth and form of planetary seedlings: results from a sounding rocket microgravity aggregation experiment. *Phys. Rev. Lett.* 93: 021103.

Kwok, S. 2004. The synthesis of organic and inorganic compounds in evolved stars. Nature 430: 985-99l.

Lamour, S., Pallmann, S., Hass, M., and Trapp, O. 2019. Prebiotic Sugar Formation Under Nonaqueous Conditions and Mechanochemical Acceleration. Life 9:52 (11 pp).

Lherminier, S., Planet R., Vehel, V. L., Simon, G., Vanel, L., Måløy, K. J., and Ramos, O. Continuously sheared granular matter reproduces in detail seismicity laws. 2019. Phys. Rev. Lett. 122: 218501.

Li, A. and Greenberg, J. M. 1997. A unified model of interstellar dust. Astron. Astrophys. 323: 566-584.

LibreTexts. 2020. www.bio.libretexts.org, Enantiomers. Last updated Aug. 15. 2020.

McClendon JH. 1976. Elemental abundance as a factor in the origins of mineral nutrient requirements. J Mol Evol 8:175-195.

McGuire, B. A. and Carroll, P. B. 2016. Mirror asymmetry in life and in space. Physics Today 69(11): 86, doi: 10.1063/PT.3.3375.

McGuire, B. A., Carroll, P. B., Loomis, R. A., Finneran, I. A., Jewell, P. R., Remijan, A. J., Blake, G. A. 2016. Discovery of the interstellar chiral molecule propylene oxide ($CH_3CHCH_2O$). Science 352: 1449-1452.

Modica, P, Meinert, C., De Marcellus, P., Nahon, L., Meierhenrich, U.J., Le Sergeant D'hendecourt, L. 2014. Enantiomeric excesses induced in amino acids by ultraviolet circularly polarized light irradiation of extraterrestrial ice analogs: A possible source of asymmetry for prebiotic chemistry. The Astrophysical Journal 788(1):79

Muller, H. J. 1927. Artificial Transmutation of the Gene. Science 66: 84–87.

Oba, Y., Takano, Y., Naraoka, H. et al. 2019. Nucleobase synthesis in interstellar ices. Nature Communication 10: 4413. https://doi.org/10.1038/s41467-019-12404-1.

Obhodas, J, Valković, V., Kollar, R., Hrenović, J., Nađ, K., Vinković, A., Orlić, Ž. 2020. The growth and sporulation of *Bacillus subtilis* in nanotesla magnetic fields. Accepted for publishing in Astrobiology.

Okada, H. et al. 2008. Absolute polarimetry at RHIC. arXiv:0712.1389v2 [nucl-ex] 9 Jan 2008.

Ossenkopf, V. 1993. Dust coagulation in dense molecular clouds—the formation of fluffy aggregates. *Astron. Astrophys.* 280: 617–646.

Pizzarello S. 2006. The chemistry of life's origin: A carbonaceous meteorite perspective. Acc. Chem. Res. 39: 231–237.

Rafelski et al. (2012). Metallicity evolution of damped Ly-α systems out to z ≈ 5. The Astrophysical Journal 755: 89 (21 pp).

Rafelski et al. (2014). The rapid decline in metallicity of damped Lyα systems at z∼5. The Astrophysical Journal Letters 782: L29 (6pp).

Reeves, H. 1994. On the origin of the light elements (Z<6). Revs. Modern Physics 66: 193-216.

Rémy-Ruyer, A., Madden, S. C., Galliano, F. 1994. Gas-to-dust mass ratios in local galaxies over a 2 dex metallicity range. Astronomy and Astrophysics 563: A31 (p 22).

Sabri, T. et al. 2014. Interstellar silicate analogs for grain-surface reaction experiments: gas-phase condensation and characterization of the silicate dust grains. *Astrophys. J*. 780: 180.

Scorei R. 2012. Is Boron a Prebiotic Element? A Mini-review of the Essentiality of Boron for the Appearance of Life on Earth. Orig Life Evol Biosph; 42(1):3-17. doi: 10.1007/s11084-012-9269-2.

Sczepanski, J. T. and Joyce, G. F. 2014. A cross-chiral RNA polymerase ribozyme. Nature 515(7527): 440-442.

Stolar, T. et al. 2020. DNA-specific selectivity in pairing of model nucleobases in the solid state. ChemComm, The Royal Society of Chemistry, Communication (4 pp).

Tarduno JA, Cottrell RD, Bono RK, Oda H, Davis WJ, Fayek M, Van't Erve O, Nimmo F, Huang W, Thern ER, Fearn S, Mitra G, Smirnov AV, Blackman EG. 2020. Paleomagnetism indicates that primary magnetite in zircon records a strong Hadean geodynamo. Proceedings of the National Academy of Sciences. 117(5):2309-2318. doi: 10.1073/pnas.1916553117.

Tazaki, R., Tanaka, H., Okuzumi, S., Kataoka, A. and Nomura, H. 2016. Light scattering by fractal dust aggregates. I. Angular dependence of scattering. *Astrophys. J*. 823: 70.

The Cologne Database for Molecular Spectroscopy, http://www.astro.uni-koeln.de/cdms/molecules.

van Dishoeck, E. F. 2014. Astrochemistry of dust, ice and gas: introduction and overview. *Faraday Discuss.* 168: 9–47.

Weiss, M. C., Sousa, F. L., Mrnjavac, N., Neukirchen, S., Roettger, M., Nelson-Sathi, S. and Martin, W. F. 2016. The physiology and habitat of the last universal common ancestor. Nature Microbiology 1, Article number: 16116.

Whittet, D. C. B. 2003. Dust in the Galactic Environment. IOP Publishing, Bristoil, UK.

Wurm, G. and Blum, J. 1998. Experiments on preplanetary dust aggregation. *Icarus* 132: 125–136.

Yin, P., Zhang, Z., Lv, H. et al. 2015. Chiral recognition and selection during the self-assembly process of protein-mimic macroanions. Nature Communication 6: 6475. https://doi.org/10.1038/ncomms7475

Zel'dovich, B., Saakyan, D., and Sobel'man, I. 1977. Energy difference between right-hand and left-hand molecules, due to parity nonconservation in weak interactions of electrons with nuclei. JETP Letters, 25(2): 94.